\documentclass[aj,twocolumn]{aastex62}

\usepackage{placeins}
\usepackage{comment}
\newcommand{\pllambda}{\ensuremath{2.5 _{- 0.9 }^{+ 1.0 }}}

\newcommand{\ctbd}[1]{}

\newcommand{\Lc}{Light curve}

\newcommand{\kms}{\ensuremath{\rm km\,s^{-1}}}

\newcommand{\rstar}{\ensuremath{R_\star}}

\newcommand{\rpl}{\ensuremath{R_{p}}}

\newcommand{\arstar}{\ensuremath{a/\rstar}}

\graphicspath{{./}{figures/}}

\accepted{\today}
\submitjournal{ApJL}

\shorttitle{Obliquity of DS Tuc Ab}
\shortauthors{Zhou et al.}
\begin{document}

\title{A well aligned orbit for the 45 Myr old transiting Neptune DS Tuc Ab}

\correspondingauthor{George~Zhou}
\email{george.zhou@cfa.harvard.edu}

\author[0000-0002-4891-3517]{G.~Zhou}
\affiliation{Center for Astrophysics \textbar{} Harvard \& Smithsonian, 60 Garden St., Cambridge, MA 02138, USA.}
\affiliation{Hubble Fellow}

\author[0000-0002-4265-047X]{J.N.~Winn}
\affiliation{Department of Astrophysical Sciences, Princeton University, NJ 08544, USA.}

\author[0000-0003-4150-841X]{E.R.~Newton}
\affiliation{Department of Physics and Astronomy, Dartmouth College, Hanover, NH 03755, USA}

\author[0000-0002-8964-8377]{S.N.~Quinn}
\affiliation{Center for Astrophysics \textbar{} Harvard \& Smithsonian, 60 Garden St., Cambridge, MA 02138, USA.}

\author[0000-0001-8812-0565]{J.E.~Rodriguez}
\affiliation{Center for Astrophysics \textbar{} Harvard \& Smithsonian, 60 Garden St., Cambridge, MA 02138, USA.}

\author[0000-0003-3654-1602]{A.W.~Mann}
\affiliation{Department of Physics and Astronomy, The University of North Carolina at Chapel Hill, Chapel Hill, NC 27599, USA}

\author[0000-0001-9982-1332]{A.C.~Rizzuto}
\affiliation{Department of Astronomy, The University of Texas at Austin, Austin, TX 78712, USA}
\affiliation{51 Pegasi b Fellow}

\author[0000-0001-7246-5438]{A.M.~Vanderburg}
\affiliation{Department of Astronomy, The University of Texas at Austin, Austin, TX 78712, USA}
\affiliation{Sagan Fellow}

\author[0000-0003-0918-7484]{C.X.~Huang}
\affiliation{Department of Physics, and Kavli Institute for Astrophysics and Space Research, Massachusetts Institute of Technology, Cambridge, MA 02139, USA.}

\author[0000-0001-9911-7388]{D.W.~Latham}
\affiliation{Center for Astrophysics \textbar{} Harvard \& Smithsonian, 60 Garden St., Cambridge, MA 02138, USA.}

\author{J.K.~Teske} %PFS observers
\affiliation{The Observatories of the Carnegie Institution for Science, 813 Santa Barbara St., Pasadena, CA 91101, USA}
\affiliation{Hubble Fellow}

\author{S. Wang} %PFS observers
\affiliation{Department of Terrestrial Magnetism, Carnegie Institution for Science, 5241 Broad Branch Road, NW, Washington, DC 20015, USA}
\affiliation{The Observatories of the Carnegie Institution for Science, 813 Santa Barbara St., Pasadena, CA 91101, USA}

\author{S.A.~Shectman} %PFS observers
\affiliation{The Observatories of the Carnegie Institution for Science, 813 Santa Barbara St., Pasadena, CA 91101, USA}

\author[0000-0003-1305-3761]{R.P.~Butler} %PFS observers
\affiliation{Department of Terrestrial Magnetism, Carnegie Institution for Science, 5241 Broad Branch Road, NW, Washington, DC 20015, USA}

\author[0000-0002-5226-787X]{J.D.~Crane} %PFS observers
\affiliation{The Observatories of the Carnegie Institution for Science, 813 Santa Barbara St., Pasadena, CA 91101, USA}

\author{I. Thompson} %PFS observers
\affiliation{The Observatories of the Carnegie Institution for Science, 813 Santa Barbara St., Pasadena, CA 91101, USA}

\author{T.J.~Henry}  %CHIRON
\affiliation{RECONS Institute, Chambersburg, PA 17201, USA}

\author{L.A.~Paredes}  %CHIRON
\affiliation{Department of Physics and Astronomy, Georgia State University, Atlanta, GA 30302, USA}

\author[0000-0003-0193-2187]{W.C.~Jao}  %CHIRON
\affiliation{Department of Physics and Astronomy, Georgia State University, Atlanta, GA 30302, USA}

\author{H.S.~James}  %CHIRON
\affiliation{Department of Physics and Astronomy, Georgia State University, Atlanta, GA 30302, USA}

\author{R.~Hinojosa}  %CHIRON
\affiliation{Cerro Tololo Inter-American Observatory, CTIO/AURA Inc., La Serena, Chile}

% %% Note that the \and command from previous versions of AASTeX is now
% %% depreciated in this version as it is no longer necessary. AASTeX 
% %% automatically takes care of all commas and "and"s between authors names.

% %% AASTeX 6.2 has the new \collaboration and \nocollaboration commands to
% %% provide the collaboration status of a group of authors. These commands 
% %% can be used either before or after the list of corresponding authors. The
% %% argument for \collaboration is the collaboration identifier. Authors are
% %% encouraged to surround collaboration identifiers with ()s. The 
% %% \nocollaboration command takes no argument and exists to indicate that
% %% the nearby authors are not part of surrounding collaborations.

% %% Mark off the abstract in the ``abstract'' environment. 

\begin{abstract}

DS Tuc Ab is a Neptune-sized planet that orbits around a G star in the 45 Myr old Tucana-Horologium moving group.  Here, we report the measurement of the sky-projected
angle between the stellar spin axis and the planet's orbital axis,
based on the observation of the Rossiter-McLaughlin effect during three
separate planetary transits.
The orbit appears to be well aligned with the equator of the host star, with a projected obliquity of $\lambda = \pllambda{}\, {}^\circ$. 
%This measurement was made via three separate transit observations that mapped the spectroscopic line profile variations induced by the transiting planet on the rotating host star.   % JNW: Seems unnecessary.
In addition to the distortions in the stellar absorption lines due
to the transiting planet, we observed variations that we attribute
to large starspots, with angular sizes of tens of degrees.  
The technique we have developed for simultaneous modeling
of starspots and the planet-induced distortions may be
useful in other observations of planets around active stars. 

\end{abstract}

%% Keywords should appear after the \end{abstract} command. 
%% See the online documentation for the full list of available subject
%% keywords and the rules for their use.

%% From the front matter, we move on to the body of the paper.
%% Sections are demarcated by \section and \subsection, respectively.
%% Observe the use of the LaTeX \label
%% command after the \subsection to give a symbolic KEY to the
%% subsection for cross-referencing in a \ref command.
%% You can use LaTeX's \ref and \label commands to keep track of
%% cross-references to sections, equations, tables, and figures.
%% That way, if you change the order of any elements, LaTeX will
%% automatically renumber them.
%%
%% We recommend that authors also use the natbib \citep
%% and \citet commands to identify citations.  The citations are
%% tied to the reference list via symbolic KEYs. The KEY corresponds
%% to the KEY in the \bibitem in the reference list below. 

% #####################################################################
%% Introduction
\section{Introduction}
\label{sec:introduction}

Studying planets over a wide range of ages is the next best thing
to being able to study planet formation and evolution in real time.
The prospects for these types of studies have been greatly enhanced
by recent discoveries of transiting planets around young stars.
Data from the NASA \emph{K2} mission have been used to find
planets in the 10-Myr-old Upper-Sco moving group \citep{2016AJ....152...61M,2016Natur.534..658D}, the 20-Myr-old Taurus-Auriga group \citep{2019AJ....158...79D}, and older clusters such as Praesepe and the Hyades \citep[e.g][]{2016ApJ...818...46M, 2017AJ....153...64M,2018AJ....156...46V, 2018AJ....156..195R}.
These efforts have also resulted in the first determinations of
the occurrence rates of close-in small planets in young associations and clusters, and allowed meaningful comparisons with the planet population around mature main-sequence field stars \citep{2017AJ....154..224R}. 

DS Tuc Ab holds special importance amongst this population of planets. Discovered via observations with the \emph{Transiting Exoplanet Survey Satellite} \citep[TESS, ][]{2016SPIE.9904E..2BR}, DS Tuc Ab is a $\sim 6 \,R_\oplus$ planet residing in an 8-day period orbit around a member of the 45-Myr-old Tucana-Horologium moving group \citep{2019ApJ...880L..17N}.  What sets DS\,Tuc\,Ab apart from
most of the previously discovered planets around young stars is the exceptional
brightness of the host star $(V_\mathrm{mag} = 8.5)$.  This enables a more in-depth characterization of the system, including
the observations described here.

In this Letter, we present a determination of the sky-projected stellar obliquity of DS Tuc Ab based on ground-based optical spectroscopy spanning 3 transits.
The obliquity angle is a tracer for any orbit-misaligning dynamical processes that the
system might have undergone, assuming the initial condition
exhibited good alignment. Dynamical interactions within stellar binary or planetary systems can determine the orbital plane inclination of close-in planets \citep[e.g.][]{2007ApJ...669.1298F,2007ApJ...670..820W}. Protoplanetary disks may also become tilted due to the presence of stellar companions \citep[e.g.][]{2012Natur.491..418B}. Reviews by \citet{2018ARA&A..56..175D} and \cite{2018haex.bookE...2T} summarize the set of mechanisms that can result in high obliquity planetary orbits, and the inferences made from the ensemble of obliquity measurements we have now amassed. Measuring the orbital obliquities of young planets is one pathway to understanding the processes that sculpt planetary systems. 
%as well as the interactions that may have sculpted protoplanetary disks during planet formation.
With these observations, as well as those reported independently by \citep{2019arXiv191203794M},
DS Tuc Ab is now the youngest planetary system for which the stellar obliquity
has been measured.

% In this paper, we present spectroscopic transit measurements of the obliquity of DS Tuc Ab \citep{2019ApJ...880L..17N}. DS Tuc Ab is a $5\,R_\oplus$ Neptune residing in an 8-day period orbit around a member of the 45\,Myr Tucana-Horologium moving group. With a brightness of $V=8.5$, it is one of the first planet around a young moving group member that allows for transit spectroscopic follow-up observations. We present three transits obtained with a series of spectroscopic facilities that map out projected obliquity of the system, making DS Tuc Ab the youngest planetary system to have its orbital obliquity mapped. 

\section{Transit spectroscopic observations}
\label{sec:obs}

The basis of the measurement technique is the Rossiter-McLaughlin effect, which utilizes
changes in the profiles of the stellar absorption lines during a planetary transit.
The changes in the stellar spectrum are sometimes analyzed as
overall shifts in the central wavelengths
of the lines. Young stars
such as DS Tuc A present special challenges because they exhibit significant photometric and spectroscopic variability, due to large spots on their stellar surfaces. Here, we found it advantageous to analyze the distorted
line profiles directly to best understand the stellar spots and their effect on our orbital obliquity measurement. Similarly, we also decided to observe three different transits for consistency checks in our obliquity derivations. 
%To mitigate any possible biases these spots may bring to the eventual obliquity measurement, multiple transit epochs were obtained for consistency checks. 
% JNW: Citation here? Spell out SMARTS?

\subsection{6.5\,m Magellan -- Planet Finder Spectrograph}
\label{sec:pfs}

We observed two transits with the Planet Finder Spectrograph \citep[PFS,][]{2010SPIE.7735E..53C} on the 6.5\,m Magellan Clay Telescope, located at Las Campanas Observatory, Chile.   PFS is a high resolution echelle spectrograph, fed via a $0.3\arcsec$ slit for our observations, yielding a spectral resolving power of $R \equiv \lambda / \Delta \lambda = 130{,}000$ over a spectral range of $3910$--$7340\,\AA$.  Typically, PFS is used for precise radial-velocity measurements
and employs an iodine gas absorption cell for wavelength calibration.
We did not use the iodine cell because we wanted to analyze the stellar absorption
line profiles without the interference of the iodine spectrum.

The observations were conducted on 2019-08-19 and 2019-10-07 UTC.
On the first night, we obtained 36 spectra starting at 03:20 and ending at 09:57 UTC.
On the second night, we obtained 33 spectra in between 23:54 and 05:57 UTC. In both
cases, the observations spanned the full transit.  The integration time
was 600\,sec per spectrum.  Wavelength solutions were determined with reference
to spectra of the Thorium-Argon hallow cathode lamp, obtained at
the beginning and the end of the night.

The stellar line profiles were derived from each spectrum via a least-squares deconvolution (LSD) over the wavelength range from 4000 to $6100\,\AA$ \citep[following][]{1997MNRAS.291..658D,2010MNRAS.407..507C}. A set of synthetic spectra from the ATLAS9 model atmospheres \citep{Castelli:2004} were used as templates for the deconvolution. We constructed a master line profile based on an average of the ensemble of observations.  
%To mitigate the effects of relatively
%dark starspots, the master line profile 
%is an average of the brightest 5\% of the observations.  
The master line profile was then subtracted
from each individual spectrum.  The residuals display features due to both
the planet and the starspots, and were analyzed further as described in 
Section~\ref{sec:analysis}.

\subsection{1.5\,m SMARTS -- CHIRON}
\label{sec:chiron}

We also observed a transit of DS Tuc Ab using the CHIRON facility \citep{2013PASP..125.1336T} on the 1.5\,m Small and Moderate Aperture Research Telescope System (SMARTS) telescope, located at Cerro Tololo Inter-American Observatory (CTIO), Chile. CHIRON is a high resolution echelle spectrograph fed from an image slicer through a fiber bundle, with a spectral resolving power of $R=80{,}000$ over the wavelength region from 4100 to $8700$\,\AA{}. A total of 21 CHIRON spectra were obtained during the 2019-08-11 transit of DS Tuc Ab, from 02:26 to 05:52 UTC, with 600\,sec of integration per exposure.
This covered the full duration of the transit.  Line profiles were derived from the CHIRON spectra via the same LSD analysis described in Section~\ref{sec:pfs}. 

Stellar atmospheric parameters, including effective temperature, surface gravity, and metallicity, were also measured from the CHIRON observations. These parameters were measured by matching the CHIRON spectra against an interpolated library of $\sim 10,000$ observed spectra classified by the Spectral Classification Pipeline \citep{2012Natur.486..375B}. We find DS Tuc A to have an effective temperature of $T_\mathrm{eff} = 5660\pm100$\,K, surface gravity of $\log g = 4.5 \pm 0.1$, and metallicity of $\mathrm{[Fe/H]} = -0.1\pm0.1$. %We adopt the spectroscopically measured effective temperature and metallicity as priors in our analysis (Section~\ref{sec:analysis}). 

\section{Analysis and Modeling}
\label{sec:analysis}

As mentioned above, modeling the spectroscopic transits of DS Tuc Ab presents an interesting challenge.  The \emph{TESS} light curve shows spot-induced photometric
variability on the order of 2\%, which is ten times larger than the
amplitude of the transit signals.  We found it necessary to fit each residual spectrum
with a model that includes the influence of star spots and the Rossiter-McLaughlin effect.

The spots are modeled as circular patches on the stellar photosphere,
each with a uniform surface brightness $S_\mathrm{spot}$ which is
darker than the surrounding photosphere.
Each spot is parameterized by its angular radius on the photosphere ($R_\mathrm{spot}$), latitude ($\alpha_\mathrm{spot}$),
and initial longitude ($\phi_\mathrm{spot}$).
The longitudes are defined such that $-90^\circ$
corresponds to a distortion at the blue extreme
of the spectral line.
% JNW: Did I get that right?
% prev txt: at the blue shifted edge of the star with respect to the observer's line of sight.
This is similar to the parameterization of the spot modeling adopted by \citet{2012A&A...545A.109B}, though we adopt to use a simple grid numerical integration over the spots to compute their effects on the stellar line profile. 
For simplicity, we assumed the stellar rotation axis is perpendicular to the line of sight,
i.e., we forfeited any attempt to constrain the stellar inclination angle from the data.
This is because such determinations generally yield degenerate posteriors,
and because the assumption $\sin I_\star=1$ is compatible
with the combination of the best estimate for the
projected stellar rotation velocity ($v\sin I_\star = 19.51\, \kms$),
the photometric rotation period (2.85 days), and the stellar radius ($1.022\,R_\odot$) \citep{2019ApJ...880L..17N}. %\textbf{The spherical geometry transformations that project our spot models to the projected stellar surface are similar to that described in \citet{2012A&A...545A.109B}.}

\begin{figure*}
    \centering
    \includegraphics[width=1.0\linewidth]{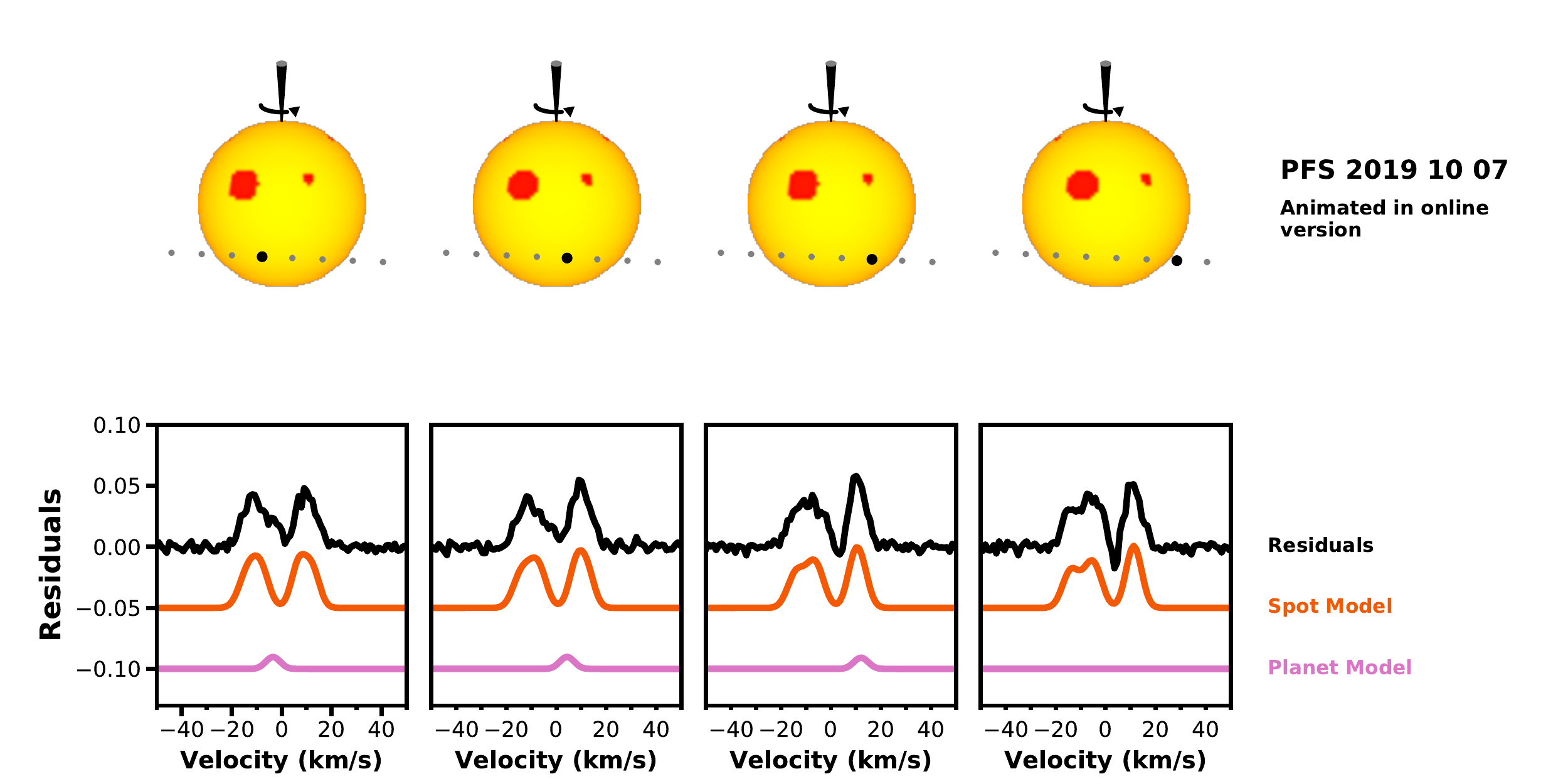}
    \caption{Line profile variations during the 2019-10-07 transit of DS Tuc Ab shown in animated form. The animation shows \textbf{(top panel)} the configuration of star spots and location of the transiting planet for our best fitting model, and \textbf{(bottom panel)} the line profile residual variations throughout the transit observation. Contributions to this signal from the star spot model (orange) and from the planet signal (purple) are shown. The animation is 5 seconds long, and shows the planet traversing through the stellar disk, and the star spots slowing rotating along with the star. The in-line still figure shows four panels of the animation, representing the variations in the line profiles seen during the transit. The last panel shows the out-of-transit line profile residual post-egress, with no contribution from the shadow of the transiting planet.  }
    % The \textbf{top panels} show the observed line profiles (black)
    % compared to the master line profile (red).
    % % JNW: Consider omitting the top panel - not very helpful.
    % The \textbf{top panels} show the locations of the starspots and the
    % transiting planet in the best-fitting model.
    % % JNW: Would be nice to make these mini-images larger.
    % The \textbf{bottom panels} show the residuals after subtracting the
    % master line profile from the observed line profile, over the four individual epochs as indicated on the top panel. Also shown are the
    % two components of the best-fitting model:
    % the effects from the starspots, and from the planet.}
    %\textbf{An animated version of this figure is available at \url{https://www.cfa.harvard.edu/~yjzhou/misc/pfs_20190819.gif} for the 2019-08-19 transit, and at \url{https://www.cfa.harvard.edu/~yjzhou/misc/pfs_20191007.gif} for the 2019-10-07 transit.}}
    \label{fig:spot_dt}
\end{figure*}

The darkening effect of each spot was calculated via a numerical
integration over the stellar surface, accounting for limb darkening,
instrumental broadening, and radial-tangential macroturbulence
\citep{2005oasp.book.....G}.
% JNW: What was the assumed level of macroturbulence? Or was it actually possible
% to fit for vmac independently of external constraints?
Another key parameter in the model is the number of spots.
We used the Bayesian Information Criterion to decide on the best
number of spots needed to fit each transit sequence.
For the 2019-08-11 CHIRON transit, we did not include
any spots in the model. 
For the 2019-08-19 and 2019-10-07 PFS transits, we used four spots. The Appendix details the Bayesian information criterion and the best fit $\lambda$ value for each trialed spot configuration in our analyses. 
Figure~\ref{fig:spot_dt} illustrates
the spot model for the 2019-10-07 transit (an animated version of both transits are available in the online edition). We note that the \emph{TESS} light curves show significant spot evolution over its observation period of $\sim 30$ days. As such we do not expect the spot configuration from our two PFS observations to be related. 

We note that these are idealistic models that likely do not capture the true complexity of the stellar surface. In particular, whilst the 2019-08-11 CHIRON transit requires no spot corrections, this is only because the average line profile subtracts well from each individual exposure given the short time baseline of the observation. Spectroscopic observations encompassing the full rotational phase coverage are required to reconstruct realistic representations of the spot coverage for DS Tuc A, as shown by classical Doppler imaging exercises \citep[e.g.][]{1997MNRAS.291..658D}. 

%That observation took place over a shorter time scale than the subsequent PFS transits,
%and the line profile variations due to spot rotation were negligible.
%For the 2019-10-07 PFS transit, we used three spots.  

We took a comprehensive approach to determining the system parameters, by performing a simultaneous global modeling with:
\begin{enumerate}
\item The PFS and CHIRON spectra spanning 3 transits,
\item The \emph{TESS} transit light curve,
\item The \emph{Spitzer} transit light curves from \citet{2019ApJ...880L..17N},
\item The spectroscopically derived stellar effective temperature and photometric spectral energy distribution available in literature,
\item The trigonometric parallax reported in Gaia Data Release 2 \citep{2018A&A...616A...1G},
\item The MIST theoretical
stellar-evolutionary models \citep{2016ApJS..222....8D,2016ApJ...823..102C}, and
\item The estimated age of the Tucana-Horologium moving group.
\end{enumerate}
Our overall approach was similar to the one described by
\citet{2019AJ....157...31Z}. We incorporate free parameters for the transit centroid timing $T_c$, period $P$, planet-star radius ratio $R_p/R_\star$, line of sight inclination $i$, projected obliquity $\lambda$, stellar mass $M_\star$ and radius $R_\star$, parallax, rotational broadening $v\sin I_\star$, macroturbulent broadening $v_\mathrm{macro}$, as well as the parameters describing each star spot described previously. Parallax and $v\sin I_\star$ are constrained by Gaussian priors about the \emph{Gaia} and spectroscopically measured values. The stellar effective temperature is also constrained by a Gaussian prior about our spectroscopic value. We fix the stellar metallicity to Solar to be consistent with the analysis in the discovery paper \citep{2019ApJ...880L..17N}. The age of the star is constrained by a Gaussian prior of $45\pm4$\,Myr, based on the age of the Tucana-Horologium moving group \citep{2014AJ....147..146K,2015MNRAS.454..593B}. Uniform priors are applied to the remaining parameters.
The light curves were modeled as per
\citet{2002ApJ...580L.171M}. 
%When fitting the spectral energy
%distribution, the local reddening was taken to be a free parameter,
%subject to an upper limit set by dust maps of
%\citet{2011ApJ...737..103S}. 
We assume no local reddening affecting the spectral energy distribution of DS Tuc Ab. 
The local maximum reddening in the region around DS Tuc A is $A(v) \sim 0.06$ from
dust maps by \citet{2011ApJ...737..103S}, and has minimal impact to our analysis.
The line profile variation induced by the
planet (its ``Doppler shadow'') was modeled by performing the
two-dimensional integral over the area of the star covered by the
planet, incorporating local radial-tangential macroturbulence, limb
darkening, and the instrument broadening. The planet's orbit was
assumed to be circular.

To determine the best-fitting parameters and the associated
uncertainties, we used the Markov Chain Monte Carlo method as
implemented in the \emph{emcee} software package
\citep{2013PASP..125..306F}. Table~\ref{tab:planetparam} gives the
results.  We also tried re-fitting the data using the same procedure
but including only the data from one of the spectroscopic transits,
rather than fitting all 3 datasets together.  We verified that the
results were consistent, regardless of which spectroscopic transit was
chosen.

\begin{figure*}
    \centering
    \textbf{PFS 2019-08-11}\\
    \includegraphics[width=0.8\linewidth]{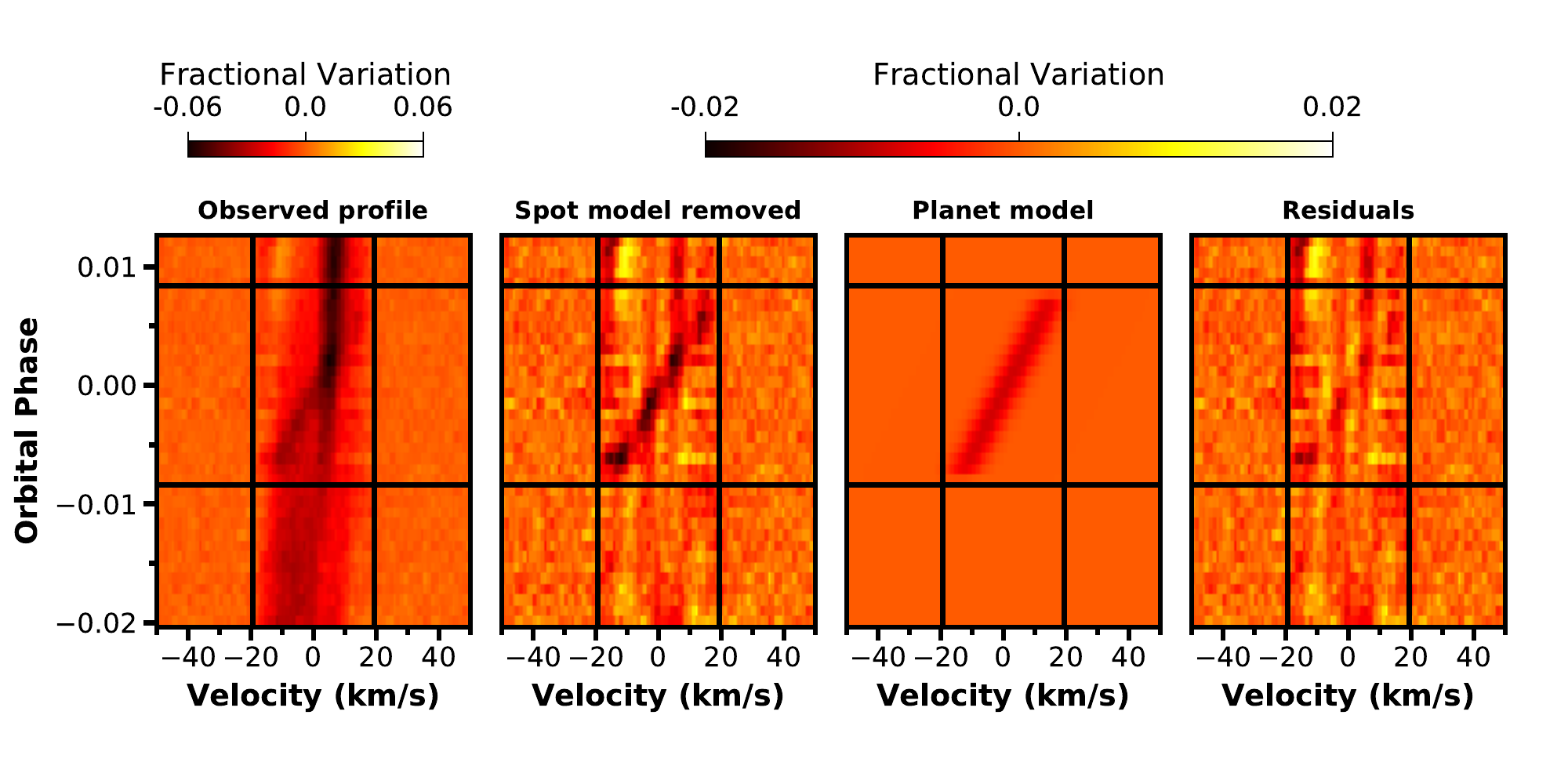}\\
    \textbf{PFS 2019-10-07}\\
    \includegraphics[width=0.8\linewidth]{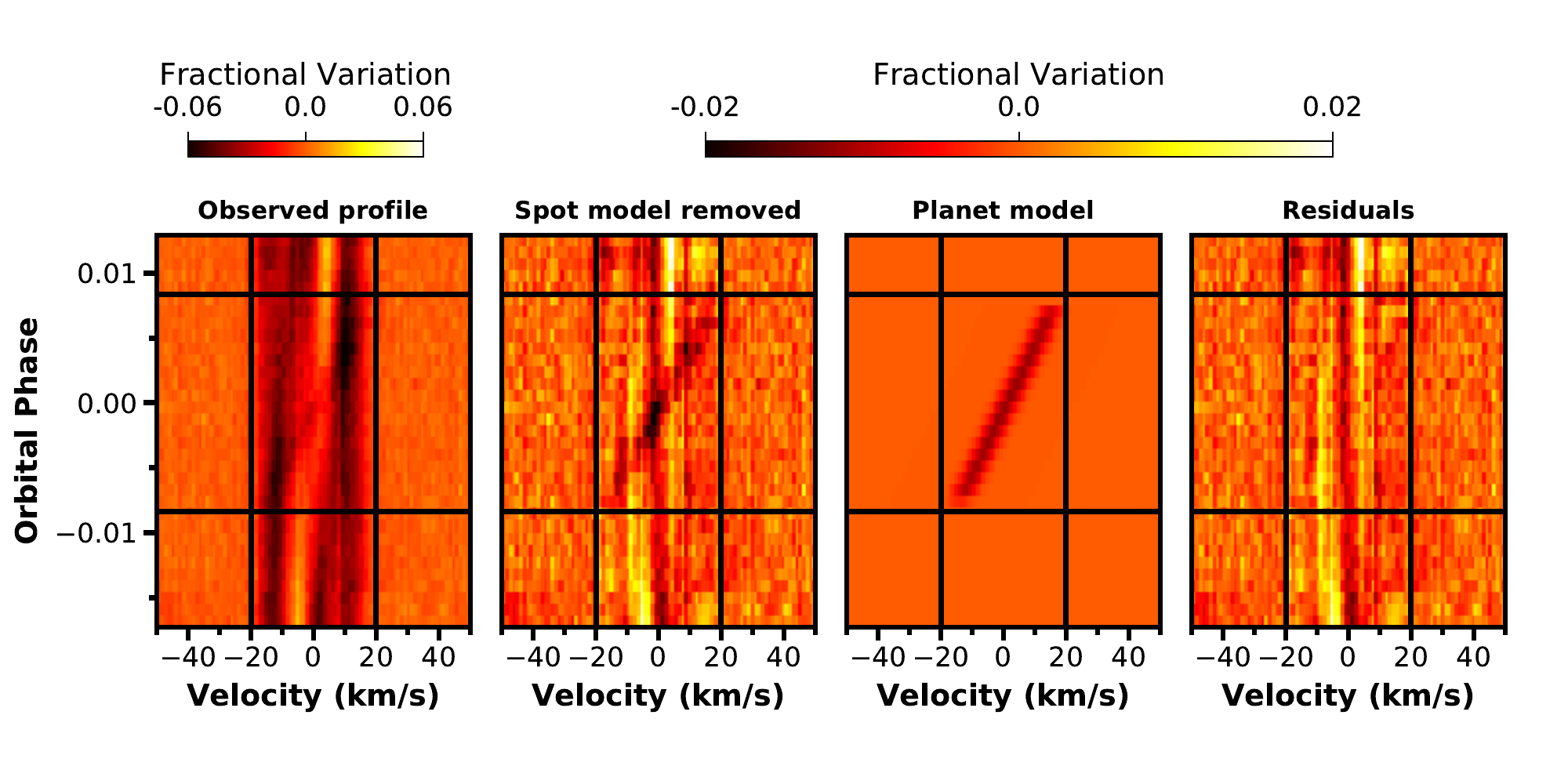}\\
    \caption{Time series of spectroscopic line profile variations observed with Magellan/PFS
    during two different transits. The horizontal axis is the velocity relative
    to the line center, the vertical axis is time (expressed as an orbital phase),
    and the color indicates the relative flux within the line.
    The horizontal lines mark the transit ingress and egress times.
    The vertical lines are drawn at $\pm v\sin I_\star$.
    The \textbf{first column} shows the line profiles
    after subtraction of the master line profile.
    The \textbf{second column} shows the residuals after further subtracting
    the model for the effects of starspots.
    The dark diagonal stripe is the planetary transit signal.
    The \textbf{third column} shows the the model for the planetary signal.
    The \textbf{fourth column} shows the residuals after subtracting the
    model for both the spots and the transit.}
    \label{fig:PFS_DT}
\end{figure*}

\begin{figure}
    \centering
    \includegraphics[width=0.8\linewidth]{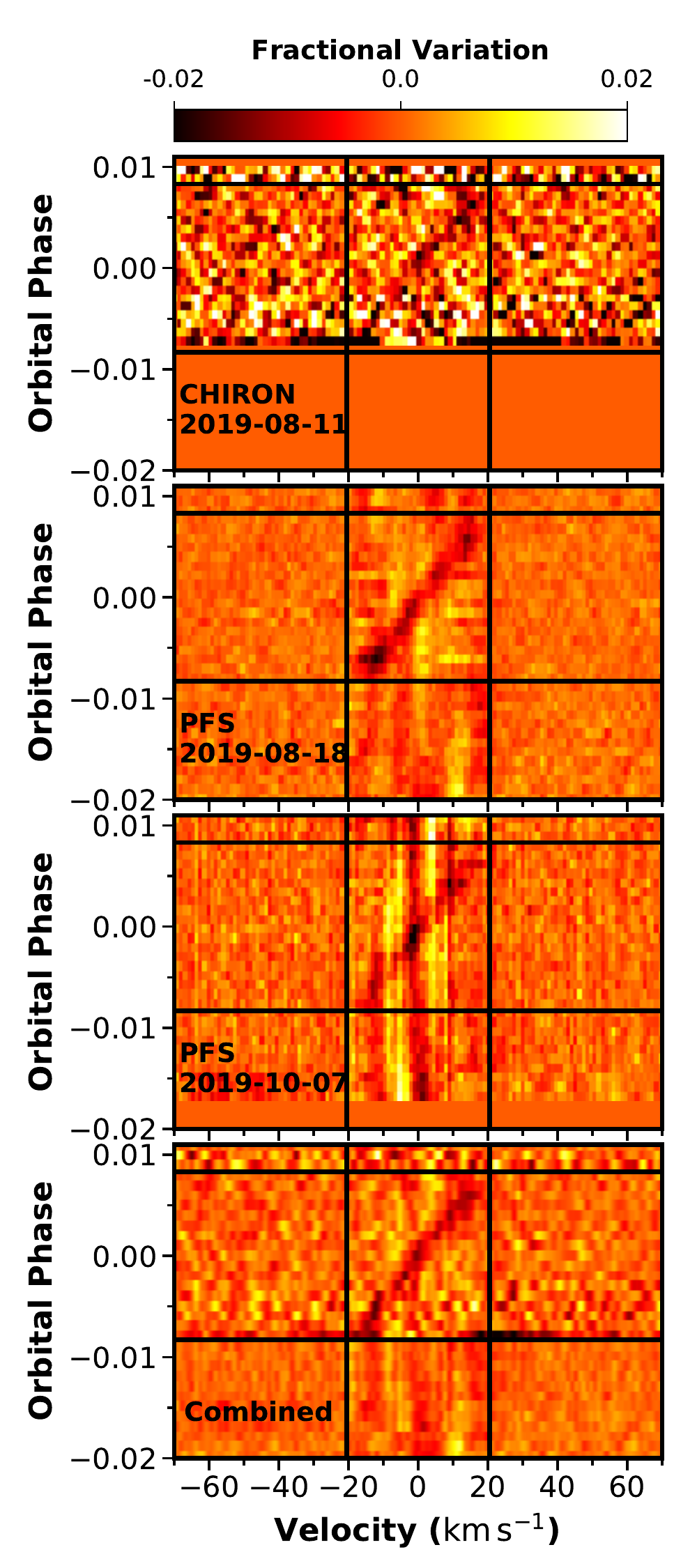}\\
    \includegraphics[width=0.7\linewidth]{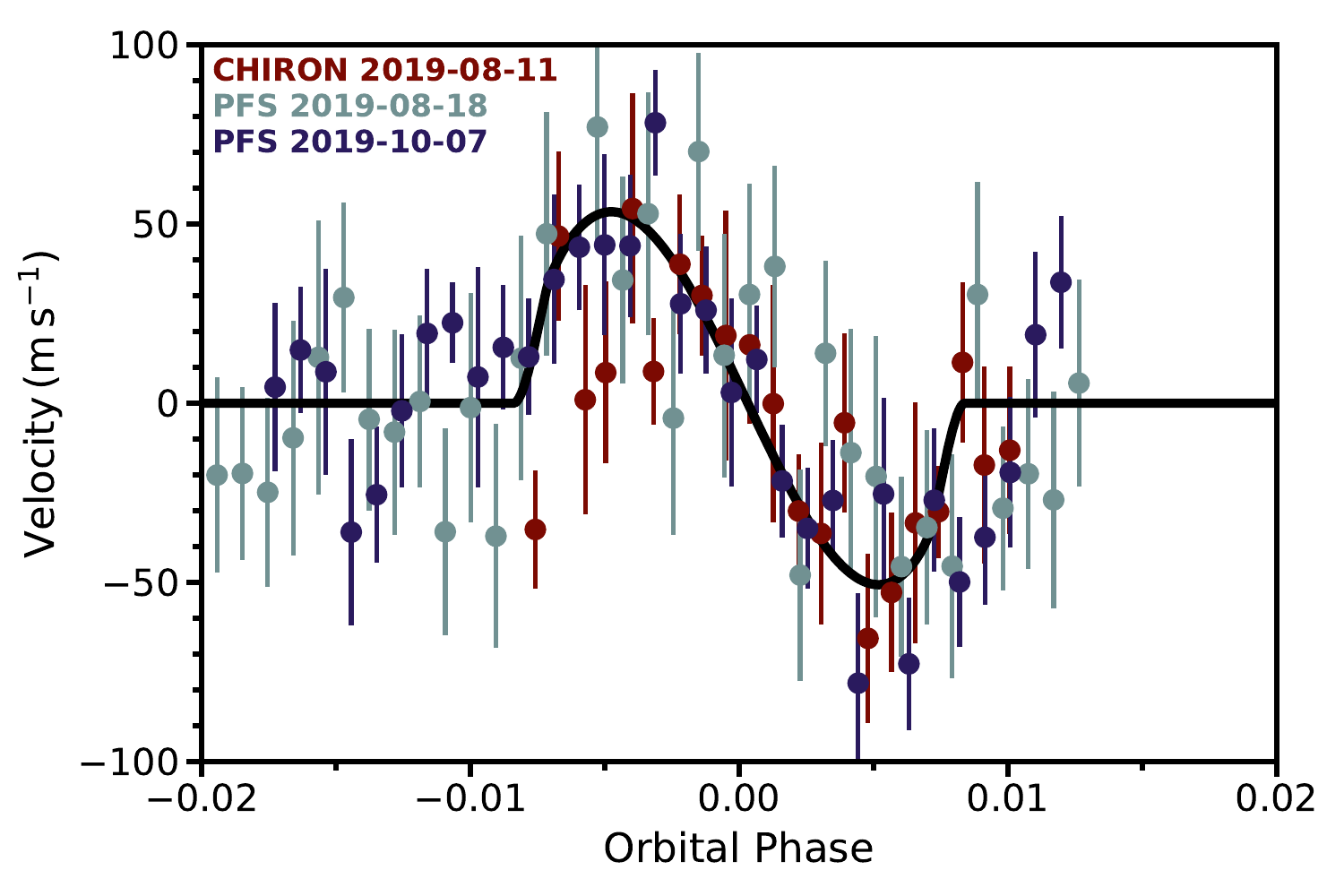}
    \caption{\textbf{Top panels} Three spectroscopic transits of DS\,Tuc\,Ab, in the same format as Figure~\ref{fig:PFS_DT}.
    The model for spot-induced variations has been subtracted.
    The planetary signal was detected in all three cases.
    The bottom panel shows the combined transit
    signal based on all three datasets. \textbf{Bottom panel} A more traditional way of displaying the Rossiter-McLaughlin effect,
    by plotting the anomalous radial velocity observed during transits.
    The black curve represents the model based on the parameters
    that were derived from our global analysis via all three transits, and is not a fit to the velocities themselves. The effects of spots have not been corrected for in these derived velocities. }
    \label{fig:dopcomb}
\end{figure}

Figure~\ref{fig:PFS_DT} depicts the analysis of the PFS spectroscopic
data observation in stages. Shown are the time series of the line
profile variations, the residuals after subtracting the model for the
effects of spots, the model for the effect of the planet, and the
residuals after subtracting all the components of the best-fitting
model.  The planetary perturbation travels the full extent of the
spectral line, from the blue side to the red side, implying that the
orbital and rotational motion are well-aligned.
Figure~\ref{fig:dopcomb} (top panels) shows all 3 spectroscopic datasets after
subtracting the best-fitting spot model, thereby isolating the
planetary signal.  The signals are all consistent with one another.

In addition, our spot model can be used to reconstruct a spot-modulated light curve of DS Tuc A. The amplitude of the model rotation curve is 10\% for  2019-08-10, and 3\% for  2019-10-07. In comparison the amplitude of the \emph{TESS} light curve spot modulation is at the 2\% level, with significant variations in the modulation amplitude through the course of the \emph{TESS} observations.

% ---------------------------------------------------------------------

\begin{deluxetable*}{lrrrr}
\tablewidth{0pc}
\tabletypesize{\scriptsize}
\tablecaption{
    Derived parameters for the DS Tuc A system
    \label{tab:planetparam}
}
\tablehead{ \\
    \multicolumn{1}{c}{~~~~~~~~Parameter~~~~~~~~}   &
    \multicolumn{1}{c}{Joint model} &
    \multicolumn{1}{c}{CHIRON 2019-08-11} &
    \multicolumn{1}{c}{PFS 2019-08-19} &
    \multicolumn{1}{c}{PFS 2019-10-07}
}
\startdata
%%
%\noalign{\vskip -3pt}
\sidehead{\textbf{\Lc{} parameters}}
~~~$P$ (days)             \dotfill    &  $ 8.138263 _{- 0.000010 }^{+ 0.000010 }$ & $ 8.138267 _{- 0.000011 }^{+ 0.000011 }$ &  $8.138264_{- 0.000011 }^{+ 0.000011 }$ & $8.138268 _{- 0.000012 }^{+ 0.000016 }$ \\
~~~$T_c$ (${\rm BJD}_{\rm TDB}$)    
  \dotfill    &  $2458332.31000 _{- 0.00024 }^{+ 0.00022 }$ &  $ 2458332.30997 _{- 0.00023 }^{+ 0.00024 }$ & $2458332.31000 _{- 0.00024 }^{+ 0.00023 }$ & $ 2458332.30995 _{- 0.00037 }^{+ 0.00026 }$\\
% ~~~$T_{14}$ (days)
%      \tablenotemark{a}   \dotfill    & $0.13342_{-0.00057}^{+0.00043}$ & $0.13398_{-0.00038}^{+0.00045}$ & $0.13488_{-0.00113}^{+0.00058}$ \\
~~~$\arstar$              \dotfill    & $16.08_{-0.28}^{+0.32}$ & $15.53_{-0.36}^{+0.40}$ & $15.73_{-0.33}^{+0.40}$ & $15.58_{-0.35}^{+0.60}$ \\
~~~$\rpl/\rstar$          \dotfill    & $ 0.05187 _{- 0.00017 }^{+ 0.00022 }$ &  $ 0.05222 _{- 0.00024 }^{+ 0.00024 }$ & $0.05216 _{- 0.00026 }^{+ 0.00024 }$ & $ 0.05219 _{- 0.00030 }^{+ 0.00025 }$ \\
~~~$i$ (deg)              \dotfill    & $ 87.76 _{- 0.10 }^{+ 0.12 }  $ & $87.56 _{- 0.13 }^{+ 0.14 }$ &  $ 87.63 _{- 0.11 }^{+ 0.12 }$ & $ 87.57 _{- 0.13 }^{+ 0.22 }$\\
\sidehead{\vspace{2mm}\textbf{Stellar parameters}}
~~~$M_\star$ ($M_\odot$)  \dotfill    &  $ 0.966 _{- 0.022 }^{+ 0.013 }$ &  $ 0.9432 _{- 0.0064 }^{+ 0.0205 }$ &  $0.9451 _{- 0.0076 }^{+ 0.0234 }$ &  $ 0.9445 _{- 0.0071 }^{+ 0.0300 }$\\
~~~$R_\star$ ($R_\odot$)  \dotfill    & $ 1.043 _{- 0.021 }^{+ 0.021 }$ &  $ 1.076 _{- 0.026 }^{+ 0.025 }$ &   $ 1.063 _{- 0.025 }^{+ 0.022 }$ & $ 1.072 _{- 0.039 }^{+ 0.026 }$\\
~~~Age (Myr)  \dotfill    &  $ 44.3 _{- 4.3 }^{+ 4.3 }$ & $ 44.83 _{- 4.94 }^{+ 4.96 }$ &  $ 44.93 _{- 3.82 }^{+ 3.98 }$ &$ 45.6 _{- 4.4 }^{+ 6.0 }$\\
%~~~$\mathrm{[Fe/H]}$ (dex) \dotfill & $ -0.124 _{- 0.057 }^{+ 0.050 }$ &  $ -0.250 _{- 0.049 }^{+ 0.026 }$ &  $ 0.076 _{- 0.026 }^{+ 0.053 }$ & $ -0.115 _{- 0.061 }^{+ 0.066 }$\\
~~~$v\sin I\star$ ($\mathrm{km\,s}^{-1}$)  \dotfill    & $ 20.58 _{- 0.24 }^{+ 0.31 }$ & $ 18.11 _{- 0.54 }^{+ 0.52 }$ &  $20.01 _{- 0.28 }^{+ 0.31 }$ & $ 19.95 _{- 0.45 }^{+ 0.31 }$\\
~~~$v_\mathrm{macro}$ ($\mathrm{km\,s}^{-1}$)  \dotfill    &  $ 2.624 _{- 0.071 }^{+ 0.064 }$ &  $ 3.44 _{- 0.60 }^{+ 0.74 }$ & $2.88 _{- 0.18 }^{+ 0.11 }$ & $ 2.24 _{- 0.14 }^{+ 0.22 }$\\
~~~$u_{1\,\mathrm{TESS}}$ Linear limb darkening coefficient  \dotfill    & 0.358 (fixed) &&\\
~~~$u_{2\,\mathrm{TESS}}$ Quadratic limb darkening coefficient  \dotfill    & 0.249 (fixed) &&\\
~~~$u_{1\,\mathrm{4.5}\mu\mathrm{m}}$ Linear limb darkening coefficient  \dotfill    & 0.0697 (fixed)&&\\
~~~$u_{2\,\mathrm{4.5}\mu\mathrm{m}}$ Quadratic limb darkening coefficient  \dotfill    & 0.1416 (fixed) &&\\
\sidehead{\vspace{2mm}\textbf{Star Spot parameters}}
~~~2019-08-19 $S_\mathrm{spot,1}$ Spot 1 Contrast  \dotfill    & $ 0.423 _{- 0.051 }^{+ 0.084 }$ & & $0.461 _{- 0.064 }^{+ 0.062 }$ \\
~~~2019-08-19 $R_\mathrm{spot,1}$ (deg) Spot 1 Radius \dotfill    &   $ 31.7 _{- 2.4 }^{+ 2.7 }$ & &  $ 30.3 _{- 2.3 }^{+ 2.5 }$\\
~~~2019-08-19 $\phi_\mathrm{spot,1}$ (deg) Spot 1 Initial Phase \dotfill    &  $ -97.2 _{- 1.9 }^{+ 4.0 }$ & &$ -93.7 _{- 1.0 }^{+ 2.7 }$ \\
~~~2019-08-19 $\alpha_\mathrm{spot,1}$ (deg) Spot 1 Latitude \dotfill    &  $ 41.4 _{- 2.7 }^{+ 3.1 }$ & & $ 40.6 _{- 2.8 }^{+ 2.4 }$\vspace{2mm}\\
%%%%%%%%%
~~~2019-08-19 $S_\mathrm{spot,2}$ Spot 2 Contrast  \dotfill    &  $ 0.463 _{- 0.045 }^{+ 0.091 }$ & &  $0.476 _{- 0.092 }^{+ 0.042 }$ \\
~~~2019-08-19 $R_\mathrm{spot,2}$ (deg) Spot 2 Radius \dotfill    &   $ 22.7 _{- 2.1 }^{+ 3.2 }$ & & $ 21.5 _{- 2.5 }^{+ 1.7 }$ \\
~~~2019-08-19 $\phi_\mathrm{spot,2}$ (deg) Spot 2 Initial Phase \dotfill    &  $ -30.5 _{- 1.8 }^{+ 1.8 }$ & & $ -32.5 _{- 2.7 }^{+ 1.5 }$\\
~~~2019-08-19 $\alpha_\mathrm{spot,2}$ (deg) Spot 2 Latitude \dotfill    &$ 4.3 _{- 7.2 }^{+ 9.0 }$ & & $ -5.4 _{- 9.4 }^{+ 10.2 }$\vspace{2mm}\\
%%%%%%%%%%
~~~2019-08-19 $S_\mathrm{spot,3}$ Spot 3 Contrast  \dotfill    &  $ 0.226 _{- 0.035 }^{+ 0.037 }$ & &  $ 0.277 _{- 0.042 }^{+ 0.052 }$ \\
~~~2019-08-19 $R_\mathrm{spot,3}$ (deg) Spot 3 Radius \dotfill    &$ 4.0 _{- 0.7 }^{+ 2.1 }$ & & $ 4.1 _{- 0.8 }^{+ 4.6 }$ \\
~~~2019-08-19 $\phi_\mathrm{spot,3}$ (deg) Spot 3 Initial Phase \dotfill    & $ -7.89 _{- 1.08 }^{+ 3.16 }$ & & $ -8.6 _{- 2.3 }^{+ 0.9 }$\\
~~~2019-08-19 $\alpha_\mathrm{spot,3}$ (deg) Spot 3 Latitude \dotfill    & $ 38.4 _{- 4.0 }^{+ 16.3 }$ & & $ 36.4 _{- 11.9 }^{+ 2.3 }$\vspace{2mm}\\
%%%%%%%%%%
~~~2019-08-19 $S_\mathrm{spot,4}$ Spot 4 Contrast  \dotfill    &  $ 0.554 _{- 0.046 }^{+ 0.066 }$ & &  $ 0.618 _{- 0.053 }^{+ 0.051 }$ \\
~~~2019-08-19 $R_\mathrm{spot,4}$ (deg) Spot 4 Radius \dotfill    &$ 20.7 _{- 1.2 }^{+ 1.5 }$ & & $ 21.8 _{- 1.1 }^{+ 2.0 }$ \\
~~~2019-08-19 $\phi_\mathrm{spot,4}$ (deg) Spot 4 Initial Phase \dotfill    & $ 27.1 _{- 2.3 }^{+ 4.2 }$ & & $ 26.8 _{- 1.9 }^{+ 2.1 }$\\
~~~2019-08-19 $\alpha_\mathrm{spot,4}$ (deg) Spot 4 Latitude \dotfill    & $ 47.7 _{- 1.2 }^{+ 1.8 }$ & & $ 46.9 _{- 1.0 }^{+ 1.2 }$\vspace{2mm}\\
%%%%%%%%%%%%%%%%%%%%%%
%%%%%%%%%%%%%%%%%%%%%%
~~~2019-10-07 $S_\mathrm{spot,1}$ Spot 1 Contrast  \dotfill    &   $ 0.573 _{- 0.047 }^{+ 0.047 }$ & & & $ 0.461 _{- 0.038 }^{+ 0.048 }$ \\
~~~2019-10-07 $R_\mathrm{spot,1}$ (deg) Spot 1 Radius \dotfill    & $ 11.2 _{- 1.6 }^{+ 1.6 }$ & & & $ 11.07 _{- 2.8 }^{+ 4.2 }$\\
~~~2019-10-07 $\phi_\mathrm{spot,1}$ (deg) Spot 1 Initial Phase \dotfill    &   $ -97.6 _{- 1.4 }^{+ 2.4 }$ & & &   $ -94.3 _{- 0.5 }^{+ 1.7 }$\\
~~~2019-10-07 $\alpha_\mathrm{spot,1}$ (deg) Spot 1 Latitude \dotfill    & $ 37.5 _{- 1.2 }^{+ 1.1 }$ & & & $ 34.1 _{- 2.8 }^{+ 1.6 }$\vspace{2mm}\\
%%%%%%%%%%
~~~2019-10-07 $S_\mathrm{spot,2}$ Spot 2 Contrast  \dotfill    & $ 0.320 _{- 0.016 }^{+ 0.016 }$ & & & $ 0.301 _{- 0.021 }^{+ 0.023 }$ \\
~~~2019-10-07 $R_\mathrm{spot,2}$ (deg) Spot 2 Radius \dotfill    &$ 11.4 _{- 1.2 }^{+ 1.1 }$ & & & $ 13.5 _{- 1.7}^{+ 1.2 }$ \\
~~~2019-10-07 $\phi_\mathrm{spot,2}$ (deg) Spot 2 Initial Phase \dotfill    & $ -44.1 _{- 0.7 }^{+ 0.6 }$ & & & $ -45.4 _{- 0.9 }^{+ 1.2 }$\\
~~~2019-10-07 $\alpha_\mathrm{spot,2}$ (deg) Spot 2 Latitude \dotfill    & $ 13.3 _{- 3.5 }^{+ 2.2 }$ & & & $ 6.7 _{- 11.0 }^{+ 8.2 }$\vspace{2mm}\\
%%%%%%%%%%
~~~2019-10-07 $S_\mathrm{spot,3}$ Spot 3 Contrast  \dotfill    & $ 0.265 _{- 0.011 }^{+ 0.012 }$ & &  & $ 0.216 _{- 0.014 }^{+ 0.047 }$ \\
~~~2019-10-07 $R_\mathrm{spot,3}$ (deg) Spot 3 Radius \dotfill    &  $ 4.0 _{- 0.7 }^{+ 1.1 }$ & &  &$ 4.7 _{- 1.1 }^{+ 1.3 }$\\
~~~2019-10-07 $\phi_\mathrm{spot,3}$ (deg) Spot 3 Initial Phase \dotfill    &   $ 3.9 _{- 0.5 }^{+ 0.5 }$ & & & $ 4.17 _{- 0.49 }^{+ 0.84 }$\\
~~~2019-10-07 $\alpha_\mathrm{spot,3}$ (deg) Spot 3 Latitude \dotfill    & $ 17.9 _{- 2.0 }^{+ 4.0 }$ & &  &$ 17.7 _{- 1.8 }^{+ 4.8 }$\vspace{2mm}\\
%%%%%%%%%%
~~~2019-10-07 $S_\mathrm{spot,4}$ Spot 4 Contrast  \dotfill    & $ 0.402 _{- 0.022 }^{+ 0.022 }$ & &  & $ 0.56 _{- 0.100 }^{+ 0.027 }$ \\
~~~2019-10-07 $R_\mathrm{spot,4}$ (deg) Spot 4 Radius \dotfill    &  $ 3.25 _{- 0.19 }^{+ 0.67 }$ & &  &$ 11.9 _{- 7.4 }^{+ 1.8 }$\\
~~~2019-10-07 $\phi_\mathrm{spot,4}$ (deg) Spot 4 Initial Phase \dotfill    &  $ 57.0 _{- 3.3 }^{+ 2.0 }$ & & & $ 68.1 _{- 2.4 }^{+ 2.1 }$\\
~~~2019-10-07 $\alpha_\mathrm{spot,4}$ (deg) Spot 4 Latitude \dotfill    & $ 51.82 _{- 1.03 }^{+ 0.79 }$ & &  &$ 48.4 _{- 1.3 }^{+ 1.2 }$\vspace{2mm}\\
\sidehead{\vspace{2mm}\textbf{Planetary parameters}}
~~~$\rpl$ ($R_\oplus$)       \dotfill    & $5.94_{-0.13}^{+0.12}$ & $6.14_{-0.17}^{+0.16}$ & $6.06_{-0.16}^{+0.15}$ & $6.11_{-0.26}^{+0.16}$\\
~~~$|\lambda|$ (deg)      \dotfill    & \pllambda{} &  $  4.2_{- 3.4 }^{+ 4.6 }$ &  $ 2.1 _{- 1.6 }^{+ 1.7 }$ &$ 2.5 _{- 1.9 }^{+ 2.0 }$ \\
~~~$a$ (AU)      \dotfill    &$0.07840 _{-0.00049}^{+0.00044}$ & $0.07763_{-0.00018}^{+0.00056}$ & $0.07768_{-0.00020}^{+0.00063}$ & $0.07766_{-0.00019}^{+0.00080}$\\
\enddata
\end{deluxetable*}

\subsection{The Rossiter-McLaughlin effect}

It is also possible to visualize the results by computing the
apparent radial velocity shift of the entire line profile induced
by all the line profile distortions.  Figure~\ref{fig:dopcomb} (bottom panel)
shows these ``anomalous radial velocities'' based on all 3 transit observations.
Before making this plot, long-term trends in the PFS radial velocities were
removed by subtracting the best-fitting quadratic function of time
to the out-of-transit data.
This trend need not be astrophysical; because the observations were obtained without the use
of an iodine gas absorption cell, the radial velocities 
are susceptible to intra-night drift of the instrumental profile and wavelength solution.
We computed a model for the anomalous radial velocity
based on the system parameters in Table~\ref{tab:planetparam}, using code provided by
\citet{2013A&A...550A..53B}.  The model agrees well with the data,
even without any further tuning of the model parameters.
With this way of displaying and analyzing the data, the effects
of spots are not as obvious. This is because the spots migrate smoothly across a significant part of the stellar surface
over the time scale of the transit. 

We note, though, that not accounting for the influence of spots can cause a systematic bias in the resulting derived obliquities from the Rossiter-McLaughlin effect. If one part of the star is dimmer due to the presence of a group of spots, then the overall weighted velocity variation induced by a planet with be affected by the specific part of the star the planet transits through. This bias is irrespective of any possible spot-crossing events. Systematic uncertainties arising from such issues will need to be accounted in the error budget in Rossiter-McLaughlin observations \citep[e.g.][]{2018A&A...619A.150O}. A light curve motivated spot-modeling exercise, in conjunction with an Rossiter-McLaughlin observation, can encompass such systematic uncertainties \citep{2019arXiv191203794M}. 

%We note that if future planetary systems orbit stars that are more rapidly rotating, significant velocity variation may occur during the transit that may not be so trivial to remove. 

%This is because the spot pattern
%is essentially static during the few hours spanning the transit.
%Future observations that make use of only the transit velocities may be
%oblivious to the complex stellar variations for active stars.

% \begin{figure}
%     \centering
%     \includegraphics[width=1.\linewidth]{RM.pdf}
%     \caption{A more traditional way of displaying the Rossiter-McLaughlin effect,
%     by plotting the anomalous radial velocity observed during transits.
%     A smoothly-varying quadratic trend in the PFS velocities was removed
%     before plotting.
%     The black curve represents the model based on the parameters
%     that were derived from the line-profile analysis.}
%     \label{fig:rmeffect}
% \end{figure}

\section{Discussion}
 \label{sec:discussion}

\begin{figure} \centering
    \includegraphics[width=1\linewidth]{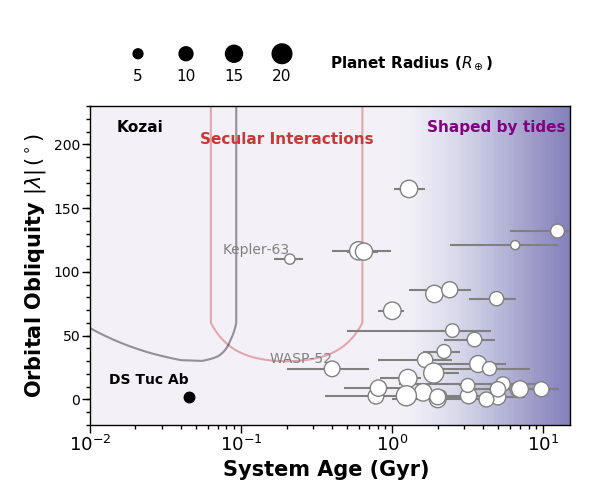} \caption{The
    age-obliquity distribution for planets with obliquity
    measurements. DS Tuc Ab is the youngest system with its orbital
    obliquity measured by nearly an order of magnitude. Other systems
    with ages $<0.5$ Gyr are labelled. The vast majority of planetary
    systems have ages estimated from isochrone modeling. Note that
    Kepler-63 (labelled) has an age estimated from gyro-chronology and
    spectroscopic activity indicators \citep{2013ApJ...775...54S}. Kozai and secular dynamical interactions can form a portion of the close-in Neptune and Jovian population. Their representative timescales and resulting orbital obliquities are noted in the figure. DS Tuc Ab is unlikely to be the result of such dynamical migration processes. Systems involving giant close-in planets older than $\sim 1$\,Gyr are likely to have experienced planet-star tidal interactions that can modify the observed obliquity distribution. Stellar ages from NASA
    Exoplanet Archive \citep{2013PASP..125..989A}, accessed 2019-09-26} \label{fig:age_lambda} \end{figure}

The finding of a well-aligned star and planetary orbit would not have
been surprising a decade ago.  Since then, though, we
have learned that stellar obliquities are sometimes very large \citep[e.g.][]{2010ApJ...718L.145W,2012ApJ...757...18A},
and no longer take spin-orbit alignment for granted. High obliquity orbits have been interpreted as evidence for
dynamical processes that tilt the orbit of a giant planet after its
formation within the gaseous protoplanetary disk. The large
majority of previous obliquity measurements have been made for close-in giant
planets around mature-age stars (Figure~\ref{fig:age_lambda}).

DS Tuc A stands out from previous systems not only because of its age. With a
radius of 6~$R_\oplus$, the planet is more representative of the
abundant population of close-orbiting planets that were revealed
clearly by the NASA Kepler mission \citep[see,
e.g.,][]{2013ApJ...778...53D,2018ApJ...860..101Z}, and have
stimulated many new ideas about planet formation and orbital
evolution.
Of these systems hosting small planets, few have had their orbital obliquities measured. Of those
with spectroscopic, spot-crossing, or astro-seismic constraints on
orbital obliquities, single-planet close-in Neptune systems can often be found in misaligned orbits, such as HAT-P-11b \citep{2010ApJ...723L.223W,2011PASJ...63S.531H,
2011ApJ...743...61S}, Kepler-63b \citep{2013ApJ...775...54S}, WASP-107b \citep{2017AJ....153..205D}, GJ436b \citep{2018Natur.553..477B}, and Kepler-408b \citep{2019AJ....157..137K}. Longer period Neptunes found in multi-planet systems, such as 
Kepler-25 \citep{2013ApJ...771...11A,2014PASJ...66...94B,2016ApJ...819...85C},
Kepler-65 \citep{2013ApJ...766..101C}, and HD 106315
\citep{2018AJ....156...93Z} have been found in well aligned orbits. Misaligned multiplanet Neptune systems do exist around main-sequence stars though \citep[HD 3167,][]{2019A&A...631A..28D}, and an extended sample of youthful planetary systems will help identify the mechanisms responsible for such systems. 

% Neptune systems in longer period orbits
% ($P\gtrsim 5\,$days) have often been found to be well aligned
% (e.g. Kepler-25
% \citealt{2013ApJ...771...11A,2014PASJ...66...94B,2016ApJ...819...85C},
% Kepler-65 \citealt{2013ApJ...766..101C}, HD 106315
% \citealt{2018AJ....156...93Z}, and WASP-166
% \citealt{2019MNRAS.488.3067H}). However, many close-in
% ($P\lesssim5\,$days) Neptunes HAT-P-11b, WASP-107b, and GJ436b occupy
% highly misaligned orbits
% \citep{2010ApJ...723L.223W,2011PASJ...63S.531H,
% 2011ApJ...743...61S,2017AJ....153..205D,2018Natur.553..477B}. 

DS Tuc Ab has similar attributes to many in the lone hot Neptune population. It lies in a close-in 8-day period orbit, and has no additional transiting companions detected. The radius of DS Tuc Ab is expected to shrink continuously as it undergoes cooling, and is likely to be of $\sim 4\,R_\oplus$ at ages $>1$\,Gyr \citep[e.g.][]{2015ApJ...808..150H}.  
A number of mechanisms can be ruled out in
influencing the dynamical history of the system. DS Tuc is a hierarchical triple
system, with the transiting Neptune orbiting DS Tuc A, and the binary
companion DS Tuc B orbiting A at a median separation of $\sim 176$ AU,
in an orbit about DS Tuc A that is likely co-planar with the orbit of
the planet \citep{2019ApJ...880L..17N}. In systems with inclined exterior companions, fast Kozai interactions
\citep[e.g.][]{2007ApJ...670..820W} can be responsible in producing
the population of close-in hot Jupiters within timescales of $10^4$ to
$10^8$ years in hierarchical triple systems. The likely co-planar
orbit of DS Tuc B with A makes Kozai interactions unlikely the cause
of DS Tuc Ab's close-in orbit. Secular planet-planet interactions may account for a portion of the close-in Neptune population. These interactions take place over timescales of hundreds of millions of years \citep{2011ApJ...735..109W}, resulting in a uniform distribution of orbital obliquities. The young age and lack of an inclined orbit suggests that DS Tuc Ab is not an example of secular planet-planet interactions. Similarly, \citet{2018AJ....155..255Y} attributed the polar orbit of HAT-P-11b to nodal precession induced by an inclined, eccentric, non-transiting outer planet.  Such precession can occur on fast timescales ($\sim 3.5$ Myr for HAT-P-11b). Long-term radial velocity or astrometric monitoring of DS Tuc A may help reveal additional non-transiting planetary companions in the system, but the low orbital obliquity we measure may already help rule out inclined planetary perturbers.  Spin-orbit misalignment via tilting of
the protoplanetary disk may occur in some binaries at time scales of
$\sim 1$ Myr, but such excitation of the disk spin-orbit angle also
requires an inclined stellar companion
\citep[e.g.][]{2012Natur.491..418B}. \citet{2019ApJ...876..127Z} suggest that some single stars may also have inner disks tilted with respect to the stellar spin axis. The probable well aligned binary
orbit of DS Tuc B and the low obliquity of DS Tuc Ab is consistent
with the lack of any large disk inclination early in its formation history. 

With an extensive set of mechanisms that can influence the evolution of
close-in planets, establishing the obliquity distribution for young planets is one
avenue that can help disentangle the histories of these planetary
systems. Establishing this distribution requires the efforts of
all-sky surveys like the new discoveries from the \emph{TESS} mission.

\acknowledgements  
Work by G.Z. is provided by NASA through Hubble Fellowship grant HST-HF2-51402.001-A awarded by the Space Telescope Science Institute, which is operated by the Association of Universities for Research in Astronomy, Inc., for NASA, under contract NAS 5-26555.
This research has made use of the NASA Exoplanet
Archive, which is operated by the California Institute of Technology,
under contract with the National Aeronautics and Space Administration
under the Exoplanet Exploration Program. Observations on the SMARTS 1.5\,m CHIRON facility were  made through the NOAO program 2019A-0004. 
Work by J.N.W.\ was supported by the Heising-Simons Foundation. 
J.K.T. acknowledges that support for this work was provided by NASA through Hubble Fellowship grant HST-HF2-51399.001 awarded by the Space Telescope Science Institute, which is operated by the Association of Universities for Research in Astronomy, Inc., for NASA, under contract NAS5-26555.
This research has made use of the NASA Exoplanet Archive, which is operated by the California Institute of Technology, under contract with the National Aeronautics and Space Administration under the Exoplanet Exploration Program. This work makes use of the Smithsonian Institution High Performance Cluster (SI/HPC).
\facility{CHIRON, Magellan ,TESS, Spitzer, Exoplanet Archive}
%\software{emcee \citep{2013PASP..125..306F}}
% #####################################################################
%% Bibliography
%\input{biblio.tex}
%\clearpage
\bibliographystyle{apj}
\bibliography{refs}

\appendix

To determine the number of spots required to best model our spectroscopic transit observations, we attempted fits of individual observations with a varying number of spots. The BIC and resulting projected obliquity $\lambda$ values of each fit are presented in Table~\ref{tab:BIC}. 

\begin{table}
    \centering
    \caption{Bayesian information criterior (BIC) and derived obliquity angles for various trialed spot configurations }
    \begin{tabular}{ccccccc}
    \hline\hline
         & \multicolumn{2}{c}{\textbf{2019-08-11 CHIRON}} & \multicolumn{2}{c}{\textbf{2019-08-19 PFS}}& \multicolumn{2}{c}{\textbf{2019-10-07 PFS}}  \\
         No. of Spots & BIC & Projected Obliquity $\lambda\,(^\circ)$ &BIC & Projected Obliquity $\lambda\,(^\circ)$ & BIC & Projected Obliquity $\lambda\,(^\circ)$ \\
         \hline
         
0	& 4880 &	4.24 &	3750&	3.20 &	6566 &	0.45\\
1	& 5845	&4.70 &	2454 &	-0.76	& 5182	& -2.57 \\
2	& 5836 &	0.31& 	1980 &	3.77 &	2163 &	0.47\\
3	&5898 &	0.89 &1852 &	3.56 &	1872 &	2.35\\
4	&5917 &	2.88 &	1647 &	3.15 &	1811 &	2.54\\
5	&&&1942 &	3.77 &	2027 &	0.72 \\
6	&&& 1724 & 3.38 & 1961 &	4.23\\         
\hline
    \end{tabular}
    \label{tab:BIC}
\end{table}

\end{document}